\documentclass[12pt]{article}
\input epsf
\usepackage{amssymb}
\usepackage{amsmath}
\makeatletter
\@addtoreset{equation}{section}
\makeatother

\def\be{\begin{equation}}
\def\ee{\end{equation}}

\def\ba{\begin{array}}
\def\ea{\end{array}}

\newcommand{\bea}{\begin{eqnarray}}
\newcommand{\eea}{\end{eqnarray}}

\newcommand{\stanford}{\it
Department of Physics, Stanford University, Stanford, CA 94305}

\newcommand{\auth}{Renata Kallosh and Andrei Linde}

\begin{document}
\hfill{}

%\hfill{hep-th/0602061}
\vskip 2cm

\vspace{24pt}

\begin{center}
{ \Large {\bf Strings, Black Holes, and Quantum Information }}

\vspace{24pt}

\auth

\vspace{10pt}

{\stanford}

\vspace{24pt}

\underline{ABSTRACT}

\end{center}

We find multiple relations between  extremal
black holes in string theory and  2- and 3-qubit systems in quantum information theory. We show that the entropy of the axion-dilaton extremal black hole is related to the concurrence of a 2-qubit state, whereas the entropy of the STU black holes, BPS as well as non-BPS, is related to the 3-tangle of a 3-qubit state.  We relate the 3-qubit  states with the string theory states with some number of D-branes. We identify a  set of ``large''  black holes with the maximally entangled GHZ-class of states and  ``small''  black holes with  separable, bipartite  and W states. We sort out  the relation between 3-qubit states,  twistors, octonions,  and  black holes. We give a simple expression for the  entropy and the area of stretched horizon of ``small'' black holes  in terms of  a norm and 2-tangles of a 3-qubit system. Finally, we show that the most general expression for the black hole and black ring entropy in N=8 supergravity/M-theory, which is  given by the famous quartic Cartan $E_{7(7)}$ invariant, can be reduced to Cayley's hyperdeterminant describing the 3-tangle of a 3-qubit state.

\vfill

\newpage

\tableofcontents 

\
%newpage

%%%%%%%%%%%%%%%

\section{Introduction}

During the last 15 years there was a significant progress in two different fields of knowledge: a description of black holes in string theory and the theory of quantum information and quantum computing. At the first glance these two subjects may seem quite distant from each other. However, there are some general themes, such as entanglement, information and entropy, which repeatedly appear both in the theory of black holes and in the theory of quantum information.  

Studies of stringy black holes began with a discovery of a broad class of new extremal black hole solutions 
\cite{GibbMaeda}, investigation of their supersymmetry \cite{Kallosh:1992ii}, a discovery of the black hole attractor mechanism \cite{Ferrara:1995ih}, and the microscopic calculation of black hole entropy \cite{Strominger:1996sh}. Investigation of stringy black holes resulted in a better understanding of the information loss paradox in the theory of black holes, revealed nonperturbative symmetries between different  versions of string theory, and stimulated what is now called ``the second string theory revolution''  \cite{Hull:1994ys,Witten:1995ex,Polchinski:1995mt}. For  reviews on stringy black holes see \cite{bhrev}. On the other hand, there were many exciting developments in the theory of quantum computation, quantum cryptography, quantum cloning, quantum teleportation, classification of entangled states and investigation of a measure of  entanglement in the context of the quantum information theory; for a review see e.g.  \cite{Zeilinger}.
It would be quite useful to find some links between these different  sets of results.

 One of the first steps in this direction was made in a recent paper by Michael Duff \cite{Duff:2006uz}. He discovered that a complicated expression for the entropy of the so-called extremal STU black holes\footnote{The  explicit construction of  BPS black holes 
with four-charges and a finite area of the  horizon within  D=4 N=4 toroidally compactified string theory was obtained in \cite{CT1}.  
This solution has an embedding as a generating 
solution in the STU model.} obtained in \cite{Behrndt:1996hu} can be represented in a very compact way as Cayley's hyperdeterminant \cite{Cayley}, which appears in the theory of quantum information in the calculation of the measure of entanglement of the 3-qubit system (3-tangle)  \cite{Coffman,Miyake}. The STU black holes represent a broad class of classical solutions of the effective supergravity derived from string theory in \cite{Duff:1995sm}.

As emphasized in \cite{Duff:2006uz}, the intriguing relation between STU extremal black holes and 3-qubit systems in quantum information theory may be coincidental. 
It may be explained, e.g., by the fact that both theories have the same  underlying  symmetry. At the level of classical supergravity the symmetry of extremal STU black holes is $[SL(2,\mathbb{R})]^3$. This symmetry may be broken down  to  $[SL(2,\mathbb{Z})]^3$ by quantum corrections or by the requirement that the electric and magnetic charges have to be quantized. In string theory a consistent microscopic description of the extremal black holes requires $[SL(2,\mathbb{Z})]^3$ symmetry. In ABC system the symmetry is $[SL(2,\mathbb{C})]^3$.

But even if the relation between the STU black holes and the 3-qubit system boils down to their underlying symmetry, this fact by itself can be quite useful. It may allow us to obtain new classes of black hole solutions and provide their interpretation based on the general formalism of quantum information. It may also provide us  with an extremely nontrivial playground for testing the general ideas of the theory of quantum information.  It would be very interesting to see how the puzzles and  paradoxes associated with black holes may be  related to the puzzles and paradoxes of the quantum information theory.

In this paper we will pursue a detailed analysis of the  relations between the structures which appear in the theory of extremal black holes and in the theory of quantum information. 
In Section \ref{qubitss} we will describe some basic facts about  general 2- and 3-qubit systems (for the hep-th reader unfamiliar with these concepts). In Section \ref{stubh}  we will discuss the relation between the 2-qubit systems \cite{Wootters:1997id} and the axion-dilaton black holes of \cite{Kallosh:1993yg}.  We will also describe the relation between the 3-qubit systems and STU black holes  represented as string theory states  with some number of D0, D2, D4 and D6 branes. This description is known to provide a microscopic entropy via counting of states of string theory \cite{Klebanov:1996mh,Maldacena:1997de}.  This microscopic entropy  coincides with the macroscopic Hawking-Bekenstein entropy (quarter of the area of the horizon) of the STU black holes at large values of charges/branes.  Section \ref{twist} gives a dictionary between   a particular $(S|TU)$ basis of STU black holes and the twistor geometry used in the description of the 3-qubit system in \cite{Levay}. In Section \ref{class} we find a one-to-one correspondence between the states of 3-qubit systems classified in \cite{Dur}  and black holes in string theory.   In Section \ref{small} we observe an intriguing relation between the value of the subsystem entanglement and the value of the quantum corrected entropy of the so-called ``small'' black holes. These black holes in a classical approximation have zero entropy and  a singular horizon, but acquire a non-zero entropy and horizon area after quantum corrections  \cite{Dabholkar:2004yr,Dabholkar:2004dq,Sinha:2006yy}.  We give a simple expression for the  entropy of ``small'' black holes  in terms of  2-tangles of a 3-qubit system and its norm. Finally, in Section \ref{e77} we show that not only the entropy of the STU black holes, but the most general expression for the black hole and black ring entropy in N=8 supergravity/M-theory,  given by the famous Cartan $E_{7(7)}$ invariant \cite{Kallosh:1996uy}, can also be represented as Cayley's hyperdeterminant describing the 3-tangle of a 3-qubit state. This, in turn, provides a natural link between the 3-qubit states and octonions.

%%%%%%%%%%%%%%%%%%%%%

\section{Qubits and a measure of entanglement}\label{qubitss}

Let us bring up several most important definitions from quantum information theory, which will be required to understand 
the correspondence between  the language of string theory black holes and the language of the quantum information theory.

Quantum entanglement is a quantum mechanical phenomenon in which the quantum states of two or more objects have to be described with reference to each other. 

A quantum bit, or {\it qubit}  is a smallest unit of quantum information. That information is described by a state in a 2-level quantum mechanical system. The two basis states  are conventionally written as $|0 \rangle$ and $|1 \rangle$. A pure qubit state is a linear quantum superposition of those two states. This means that each qubit can be represented as a linear combination of $|0 \rangle$ and $|1 \rangle$:
\be
    | \Psi \rangle = \psi_0 |0 \rangle + \psi_1 |1 \rangle\, ,
\ee
where $\psi_0$ and $\psi_0$ are complex probability amplitudes of finding the system in a particular state when one makes measurements. This leads to a normalization condition
\be
|\Psi|^{2} = \langle \Psi| \Psi \rangle =\sum_i |\psi_i|^2 =|\psi_0|^2+|\psi_1|^2=1 \ .
\label{norm1}\ee
A 1-qubit system usually goes by the name A (Alice).

For a {\it 2-qubit} state AB (Alice and Bob) one has
 \be
    | \Psi \rangle = \psi_{00} |00 \rangle + \psi_{01} |01 \rangle +\psi_{10} |10 \rangle
    +\psi_{11} |11 \rangle \, ,
\ee  
with the corresponding normalization condition, $ \langle \Psi| \Psi \rangle =1$.
One can introduce a partial density matrix, a trace over the subsystem A,  $\rho_A={\rm Tr}_B  |\psi\rangle \langle \psi|$, and the same for B. For a pure state,  entanglement $E$ is defined as the entropy of either of the two subsystems
\be
E(\psi)= -\rm Tr(\rho_A \log_2 \rho_A)= -\rm Tr(\rho_B \log_2 \rho_B) \ .
\ee
This is von Neumann entropy of a quantum state. The properties of an AB system  are also determined by the so-called  concurrence ${\cal C}$, which is a measure of the entanglement.
Concurrence of the 2-qubit AB system in a pure state can  be given as
\be
{\cal C}= {\cal C}_{AB}= 2\sqrt {\rm det\, \rho_A}= 2\sqrt {\rm det\, \rho_B}=2|\rm det \, \psi| \ .
\label{2qubitC}\ee
These two measures of entanglement are related to each other  \cite{Wootters:1997id}:
\begin{equation}
E(C(\psi)) = - \frac{1+\sqrt{1-C^2}}{2} \log_2 \frac{1+\sqrt{1-C^2}}{2}
              - \frac{1-\sqrt{1-C^2}}{2} \log_2 \frac{1-\sqrt{1-C^2}}{2}.
\label{EofC}
\end{equation}
The function ${E}(C)$ is monotonically increasing, and
ranges from 0 to 1 as $C$ goes from 0 to 1.

For a mixed state of the AB system concurrence is  more complicated. For our purposes we will need to define  the concurrence of a particular AB state inside of a pure 3-qubit state.

\begin{figure}
\centerline{ \epsfxsize 3.7 in\epsfbox{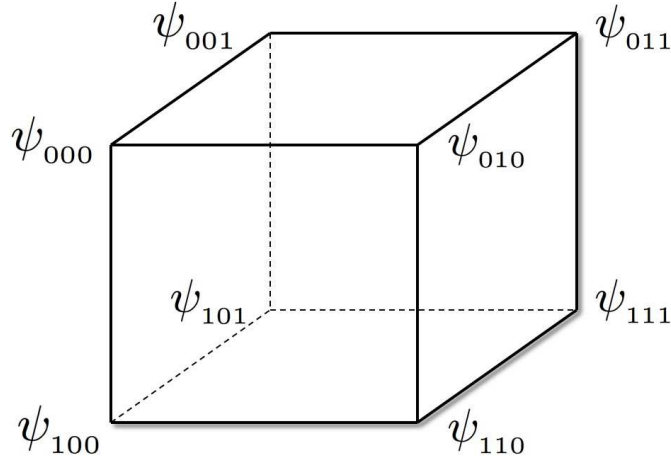}}
\caption{The $2\times 2\times 2$ matrix corresponding to the quantum state (\ref{funct}). }
\label{F1}
\end{figure}

The {\it 3-qubit} system ABC (Alice, Bob and Charlie) in turn is given by the normalized wave function 
\bea
|\Psi\rangle = \sum_{ijk=1,0}\psi_{ijk}|ijk\rangle & =& \psi_{000}|000\rangle+\psi_{001}|001\rangle+\psi_{010}|010\rangle+\psi_{011}|011\rangle
\nonumber\\
&+&\psi_{100}|100\rangle+\psi_{101}|101\rangle+\psi_{110}|110\rangle+\psi_{111}|111\rangle  \ . \label{funct}
\eea
A 3-dimensional matrix corresponding to the  3-qubit system can be represented as a cube with vertices corresponding to $\psi_{ijk}$, see Fig. \ref{F1}.

The 3-qubit system $\psi_{ijk}$ has an invariant, Cayley's hyperdeterminant \cite{Cayley} defined as\footnote{In this paper we will always write the usual determinant of a matrix $\psi_{{ij}}$ as  $\rm det\, \psi$,  and   the hyperdeterminant of a matrix $\psi_{ijk}$ as $\rm Det\, \psi$.}
\bea
{\rm Det}~\psi&=&-\frac{1}{2}\epsilon^{ii^\prime}\epsilon^{jj^\prime}\epsilon^{kk^\prime}
\epsilon^{mm^\prime}\epsilon^{nn^\prime}\epsilon^{pp^\prime}\psi_{ijk}\psi_{i^\prime j^\prime m}
\psi_{npk^\prime}\psi_{n^\prime p^\prime m^\prime} \nonumber\\
&=&  \psi_{000}^2 \psi_{111}^2 + \psi_{001}^2 \psi_{110}^2 +
        \psi_{010}^2 \psi_{101}^2 + \psi_{100}^2 \psi_{011}^2  \nonumber\\
        &-&2(\psi_{000}\psi_{001}\psi_{110}\psi_{111}+\psi_{000}\psi_{010}\psi_{101}\psi_{111}
\nonumber\\
        &+& \psi_{000}\psi_{100}\psi_{011}\psi_{111}+\psi_{001}\psi_{010}\psi_{101}\psi_{110}
\nonumber\\
        &+& \psi_{001}\psi_{100}\psi_{011}\psi_{110}+\psi_{010}\psi_{100}\psi_{011}\psi_{101}) 
\nonumber\\
        &+& 4 (\psi_{000}\psi_{011}\psi_{101}\psi_{110} + \psi_{001}\psi_{010}\psi_{100}\psi_{111}) \ .
        \label{cayley}
\eea
The 3-tangle of the ABC system as shown in  \cite{Coffman} is given by 
\be\label{coffm}
\tau_{ABC}= 4\, |{\rm Det}~\psi| \ .
\ee
When the wave function is normalized, $ \langle \Psi| \Psi \rangle =1$,
the 3-tangle $\tau_{ABC}$ is also normalized to take values in the range from 0 to 1.

An important tool in describing 3-qubit states is a reduced density matrix. For example, 
\be\label{locent}
\rho_A= {\rm Tr}_{BC} |\Psi\rangle \langle \Psi| \ , \qquad S_A= 4 ~\rm det \, \rho_A\equiv \tau_{A(BC)} \ ,
\ee
where $\rho_A$ is a $2\times 2$ matrix. $S_A$  is sometimes called local entropy, it is a measure of how entangled $A$ is with the pair $(BC)$. 
The three way tangle $\tau_{ABC}$ consists of  three contributions \cite{Coffman}:
\be
\tau_{ABC}= \tau_{A(BC)} - \tau_{AB}-\tau_{AC} \ .
\label{3tangle}
\ee 

Each term in eq. (\ref{3tangle}) is a particular contraction of the 4 terms $\psi_{ijk}$ with each other and with some number of totally antisymmetric 2-component $\epsilon^{ij}$ tensor. It was shown in \cite{Coffman}  that  the first term $\tau_{A(BC)}$, which is a tangle between Alice with Bob-and-Charlie system,  is a square of the concurrence in A(BC) system: $\tau_{A(BC)}={\cal C}^2_{A(BC)}$.  
The second term, $\tau_{AB}={\cal C}^2_{AB}$, which is called a  2-tangle between Alice and Bob in the 3-cubit system ABC,  is a square of the concurrence in AB system inside the ABC,  ${\cal C}_{AB}$ will be defined below in eq. (\ref{CAB}). Finally, the third one $\tau_{AC}={\cal C}^2_{AC}$ is the 2-tangle between Alice and Charlie in ABC; it is a square of the concurrence of the AC system inside ABC,  ${\cal C}_{AC}$ will be defined below in eq. (\ref{CAC}).
Eq. (\ref{3tangle}) and its analogues obtained by permutations of A, B and  C,  can  be represented in the form \cite{Coffman}
\be
\tau_{ABC}= {\cal C}^2_{A(BC)} - {\cal C}^2_{AB}-{\cal C}^2_{AC} \ ,
\label{3tangleA}\ee 
\be
\tau_{ABC}= {\cal C}^2_{B(CA)} - {\cal C}^2_{BC}-{\cal C}^2_{BA} \ ,
\label{3tangleB}\ee 
\be
\tau_{ABC}= {\cal C}^2_{C(BA)} - {\cal C}^2_{CB}-{\cal C}^2_{CA} \ .
\label{3tangleC}\ee 
Here 
\be
{\cal C}^2_{A(BC)}=4 ~\rm det\, \rho_A\ , \qquad {\cal C}^2_{B(AC)}=4 ~\rm det \,\rho_B\ , \qquad {\cal C}^2_{C(AB)}=4 ~\rm det \, \rho_C  \ ,
\ee
is a squared concurrence between A and the pair BC,  B and the pair AC, C and the pair AB, respectively. One can also define the concurrence of AB inside ABC in terms of various combinations of $\psi_{ijk}$.
\begin{eqnarray}\label{CAB}
{\cal C}_{AB}&=& \Big(\rm \det\,\rho_C-\rm det\, \rho_A-\rm det\, \rho_B -{1\over 2} \tau_{ABC}\Big)^{1/2} \ , \\
{\cal C}_{AC}&=&\Big(\rm \det\,\rho_B-\rm det\, \rho_A-\rm det\, \rho_C -{1\over 2} \tau_{ABC}\Big)^{1/2}\ , \label{CAC} \\
{\cal C}_{BC}&=& \Big(\rm \det\,\rho_A-\rm det\, \rho_B-\rm det\, \rho_C -{1\over 2} \tau_{ABC}\Big)^{1/2}  \ .
\label{Conc}\end{eqnarray}

In eqs. (\ref{3tangle})-(\ref{Conc}) each term scales under the rescaling of $\psi_{ijk}$ homogeneously. Thus they are valid not only for the usual normalized vectors, satisfying the condition $ \langle \Psi| \Psi \rangle =1$, but also for vectors with arbitrary norm 
\be 
|\Psi| \equiv \sqrt {\langle \Psi| \Psi \rangle} \neq  1 \ .
\label {|Psi|}\ee
One may try to interpret  $|\Psi|^2 \not = 1$  as measuring a number density rather than a probability density.
One may also notice that 
\be 
|\Psi|^2 \equiv \rho \equiv  {\rm Tr}_{ABC}|\Psi\rangle \langle \Psi|
\ee
and 
\be
\rho= {\rm Tr}_{A}\rho_A = {\rm Tr}_{B}\rho_B= {\rm Tr}_{C}\rho_C \neq 1 \ .
\ee
The difference  between normalized and unnormalized vectors plays a significant role in our subsequent analysis because we are going to use the concepts of the 2-tangle and 3-tangle not for the calculation of probabilities in quantum mechanics, but for the calculation of black hole entropy, which can be much greater than 1. In what follows we will discuss general states with norm $|\Psi| \not = 1$, and in the calculations of such objects as the 3-tangle or Cayley's hyperdeterminant we will use the states $|\Psi\rangle$ (\ref{funct}) without imposing any normalization constraints on $\psi_{ijk}$.

%%%%%%%%%%%%%%%%%%

\section{Black holes in supergravity, string theory and ABC system}\label{stubh}

%%%%%%%%%%%

\subsection{Axion-dilaton extremal black holes and concurrence of a 2-qubit system}\label{axdil}
As a warm up to STU black holes-3-qubits relation we start with a simpler case of the so-called axion-dilaton black hole solutions with manifest $SL(2,Z)$-symmetry  in \cite{Kallosh:1993yg} and display their relation to a 2-qubit system. In the case of N=2 supergravity with one vector multiplet in a version without a prepotential  the double-extremal axion-dilaton black holes were constructed in \cite{Kallosh:1993yg,Kallosh:1996xa}. The double-extreme  black holes solve the attractor equations \cite{Ferrara:1995ih} for the scalars and have everywhere constant scalars. The set of electric and magnetic charges is $ (p^0, p^1, q_0, q_1)$, and the  entropy formula is given by the following $SL(2,\mathbb{Z})$-invariant expression
\be
{S\over \pi}= |p^0q_1-q_0p^1| \ .
\ee
If we identify the charges with the components of a $2\times 2$-matrix $\psi_{ij}$
\be
\left(
\begin{array}{c}
p^0\\
p^1\\
q_1\\
q_0
\end{array}
\right)=
\left(
\begin{array}{c}
\psi_{00}\\
\psi_{01}\\
\psi_{10}\\
\psi_{11}
\end{array}
\right)\, ,
\ee
the entropy formula is proportional to the concurrence of a 2-qubit system:
\be
S=\pi | \rm det \, \psi|= {\pi\over 2} {\cal C}\ ,  \qquad \psi= \left(
\begin{array}{cc}
 p^0 & p^1 \\
  q_1 & q_0
\end{array}\right)=
\left(
\begin{array}{cc}
 \psi_{00} & \psi_{01} \\
 \psi_{10} & \psi_{11}
\end{array}\right) \ .
\ee
Thus we have identified the features in the axion-dilaton black holes with some analogous features in a 2-qubit system AB in a pure state. In particular, the entropy formula for arbitrary integer charges is equal  to the concurrence  ${\cal C}$ of the 2-qubit system described by the unnormalized vector in eq. (\ref{2qubitC}). 

%%%%%%%%%%%%%%%%

\subsection{STU black holes and 3-qubits}\label{stu3}

Consider type IIA string theory compactified on a Calabi-Yau space in presence of D0, D2, D4 and D6 branes. 
The corresponding effective N=2 supergravity is described by N=2 gravitational 
multiplet and 3 vector multiplets. First we consider the simplest version of supergravity with the prepotential $F=STU$. The electric and magnetic charges of the graviphoton are denoted
by $(p^0, q_0)$, and the ones for the 3 vector multiplets are 
 $ (p^1, q_1),\, (p^2, q_2),\, (p^3,q_3)$ respectively. These supergravity charges are known to originate  from the number of D0, D2, D4 and D6 branes as follows: the number $n_{_{D0}}$ of D0 branes is $q_0$, the numbers $k_{_{D2}}$, $m_{_{D2}}$, 
$l_{_{D2}}$ of D2 branes wrapped on 3 2-cycles are $q_1, q_2, q_3$ respectively. The numbers $k_{_{D4}}$, $m_{_{D4}}$,
$l_{_{D4}}$ of D4 branes wrapped on 3 4-cycles, dual to the relevant 2 cycles are $p^1, p^2, p^3$ and the number of D6 branes is $p^0$. Negative number of branes corresponds to a positive number of anti-branes of the same kind.

\begin{figure}[h!]
\centerline{ \epsfxsize 3.5 in\epsfbox{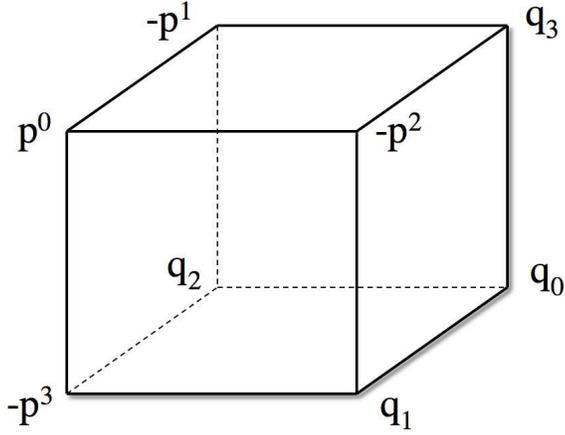}}
\caption{The $2\times 2\times 2$ matrix corresponding to supergravity black holes \cite{Behrndt:1996hu}. }
\label{stu2cube}
\end{figure}

 Following Ref. \cite{Duff:2006uz}, we can associate all magnetic charges with the presence of $1$'s in the ABC system according to a simple rule illustrated by  Eq. (\ref{dictionary}). The state with the magnetic charge $p^0$  is the state $|000\rangle$ which has zero number of $1$'s. The state with charge $p^1$ corresponds to $|001\rangle$, which has  $1$ in the first position; $p^2$ corresponds to $|010\rangle$, which has   $1$ in the second position; $p^3$ corresponds to $|100\rangle$, which has   $1$ in the third position. (We count positions from the right to the left.) We associate electric charges  with the presence of $0$'s in the ABC system. Thus the state $q_0$ corresponds to $|111\rangle$, which has no $0$'s; $q_1$, the state $|110\rangle$,  has  $0$ in the first position; $q_2$, the state $|101\rangle$,   has  $0$ in the second position; $q_3$,  the state $|011\rangle$,  has  $0$ in the third  position.  The signs are not explained by this rule, however, they have to be taken in a way so that the black hole entropy is an $[SL(2,\mathbb{Z})]^3$-invariant for integer charges and is defined by the properties of the ABC system. The explanation of signs is actually coming from the corresponding cube in $p, q$ variables given in Fig. \ref{stu2cube} which was presented in \cite{Behrndt:1996hu}.

All S-, T- and U-dualities in this basis are non-perturbative. However, one can switch to a  different basis by performing an $Sp(8, \mathbb{Z})$ transformations which transforms both the symplectic section $(X,F)$ as well as the charges $(p,q)$, e. g. 
\be
\begin{array}{c}
\left(
\begin{array}{c}
p^\Lambda\\
q_\Lambda
\end{array}
\right)'
\end{array}= \begin{array}{cc}
\left(
\begin{array}{cc}
A& B\\
C& D
\end{array}
\right)
\end{array}\begin{array}{c}
\left(
\begin{array}{c}
p^\Lambda\\
q_\Lambda
\end{array}
\right)
\end{array}
\ee
with $A^T C- C^T A= B^T D- D^T B=0$ and $A^T D- C^T B=1$.
In manifestly STU-symmetric version we have
\be
\begin{array}{c}
Supergravity\\
\\
\left(
\begin{array}{c}
p^0\\
p^1\\
p^2\\
p^3\\
q_0\\
q_{1}\\
q_{2}\\
q_{3}
\end{array}
\right)
\end{array}
=~~~
\begin{array}{c}
ABC\\
\\
\left(
\begin{array}{c}
~~ \psi_{000}\\
- \psi_{001}\\
- \psi_{010}\\
-\psi_{100}\\
~~ \psi_{111}\\
~~ \psi_{110}\\
~~ \psi_{101}\\
~~ \psi_{011}
\end{array}
\right)
\end{array}
~~~=
\begin{array}{c}
String~Theory\\
\\
\left(
\begin{array}{c}
n_{_{D6}}\\
k_{_{D4}} \\
m_{_{D4}}\\
l_{_{D4}}\\
n_{_{D0}}\\
k_{_{D2}}\\
m_{_{D2}}\\
l_{_{D2}}
\end{array}
\right)
\end{array}
\label{dictionary}\ee
It is important to stress here that the cubes, according to  Fig. \ref{F1} as well as Fig. \ref{stu2cube}  have 3 magnetic and one electric charge in  upper 4 corners, and 3 electric and one magnetic charge in lower 4 corners. 

To associate these charges/numbers of branes with the elements of the $\psi_{ijk}$ matrix one has to keep in mind that the entropy and the absolute value of the hyperdeterminant are invariant under the $[SL(2,\mathbb{Z})]^3$ subgroup of the symplectic $Sp(8, \mathbb{Z})$ transformations. 
   We may go to an $(S|TU)$ or  $(T|US)$ or $(U|ST)$ basis in which one of the duality transformations becomes perturbative and does not mix electric and magnetic charges.  In this case either S or T or U direction becomes different from other two directions.   Here are three possible options for $(p, q)'$ which one can get by returning to the symmetric $STU$ basis:
\begin{itemize}
  \item $STU \rightarrow  (S|TU)\rightarrow  STU $ \be
\begin{array}{c}
Supergravity\\
\\
\left(
\begin{array}{c}
d\, p^0+ c\, p^1\\
b\, p^{0}+ a\, p^1\\
d\, p^2 +c\, q_{3}\\
d\, p^3 + c\, q_{2}\\
a\, q_0- b\, q_{1}\\
-c\, q_0+d\, q_{1}\\
b\, p^{3}+a\, q_{2}\\
b\, p^{2}+ a\, q_{3}
\end{array}
\right)
\end{array}
=~~~
\begin{array}{c}
ABC\\
\\
\left(
\begin{array}{c}
~~ \alpha_{000}\\
- \alpha_{001}\\
- \alpha_{010}\\
-\alpha_{100}\\
~~ \alpha_{111}\\
~~\alpha_{110}\\
~~\alpha_{101}\\
~~ \alpha_{011}
\end{array}
\right)
\end{array}
~~~=
\begin{array}{c}
String~Theory\\
\\
\left(
\begin{array}{c}
d \, n_{_{D6}}+ c\,  k_{_{D4}} \\
b\, n_{_{D6}}+ a\,k_{_{D4}}\\
d\, m_{_{D4}} +c\, l_{_{D2}}\\
d\, l_{_{D4}} + c\, m_{_{D2}}\\
a\, n_{_{D0}}- b\, k_{_{D2}}\\
-c\, n_{_{D0}}+d\, k_{_{D2}}\\
b\, l_{_{D4}}+a\, m_{_{D2}}\\
b\, m_{_{D4}}+ a\, l_{_{D2}}
\end{array}
\right)
\end{array}
\label{dictionary1}\ee

  \item $STU \rightarrow  (T|US)\rightarrow  STU $
  \be
\begin{array}{c}
Supergravity\\
\\
\left(
\begin{array}{c}
d\, p^0+c\, p^{2}\\
d\, p^1+c\, q_{3}\\
b\, p_{0} + a\, p^2\\
d\, p^3+ c\, q_{1}\\
a\, q_0 - b\, q_{2}\\
b\, p^{3} + a\, q_{1}\\
-c\, q_{0} + d\, q_{2}\\
b\, p^{1}+ a\, q_{3}
\end{array}
\right)
\end{array}
=~~~
\begin{array}{c}
ABC\\
\\
\left(
\begin{array}{c}
~~ \beta_{000}\\
- \beta_{001}\\
- \beta_{010}\\
-\beta_{100}\\
~~ \beta_{111}\\
~~\beta_{110}\\
~~\beta_{101}\\
~~ \beta_{011}
\end{array}
\right)
\end{array}
~~~=
\begin{array}{c}
String~Theory\\
\\
\left(
\begin{array}{c}
d\, n_{_{D6}}+c\, m_{_{D4}}\\
d\, k_{_{D4}}+c\, l_{_{D2}}\\
b\, n_{_{D6}} + a\, m_{_{D4}}\\
d\,l_{_{D4}}+ c\, k_{_{D2}}\\
a\, n_{_{D0}} - b\, m_{_{D2}}\\
b\, l_{_{D4}} + a\, k_{_{D2}}\\
-c\, n_{_{D0}} + d\, m_{_{D2}}\\
b\, k_{_{D4}}+ a\, l_{_{D2}}
\end{array}
\right)
\end{array}
\label{dictionary2}\ee

\

  \item $STU \rightarrow  (U|ST)\rightarrow  STU $  
  \be
\begin{array}{c}
Supergravity\\
\\
\left(
\begin{array}{c}
d\, p^0+c\, p^{3}\\
d\, p^1+c\, q_{2}\\
d\, p_{2} + c\, q_{1}\\
b\, p^0+ a\, p^{3}\\
a\, q_0 - b\, q_{3}\\
b\, p^{2} + a\, q_{1}\\
b\, p^{1} + a\, q_{2}\\
-c\, q_{0}+ d\, q_{3}
\end{array}
\right)
\end{array}
=~~~
\begin{array}{c}
ABC\\
\\
\left(
\begin{array}{c}
~~ \gamma_{000}\\
- \gamma_{001}\\
- \gamma_{010}\\
-\gamma_{100}\\
~~ \gamma_{111}\\
~~\gamma_{110}\\
~~\gamma_{101}\\
~~ \gamma_{011}
\end{array}
\right)
\end{array}
~~~=
\begin{array}{c}
String~Theory\\
\\
\left(
\begin{array}{c}
d\,n_{_{D6}}+c\, l_{_{D4}}\\
d\, k_{_{D4}}+c\, m_{_{D2}}\\
d\, m_{_{D4}} + c\, k_{_{D2}}\\
b\, n_{_{D6}}+ a\, l_{_{D4}}\\
a\, n_{_{D0}} - b\, l_{_{D2}}\\
b\, m_{_{D4}} + a\, k_{_{D2}}\\
b\, k_{_{D4}} + a\, m_{_{D2}}\\
-c\, n_{_{D0}}+ d\, l_{_{D2}}\end{array}
\right)
\end{array}
\label{dictionary3}\ee
\end{itemize}

\vskip 0.5cm

According to \cite{Behrndt:1996hu}, the black hole entropy of BPS black holes is given by 
\be
{S\over \pi}= \left( W(p^\Lambda,q_\Lambda)\right)^{1/2} \ ,
\ee
where
\begin{equation} \label{ww}
 W(p^\Lambda ,q_\Lambda) =-{(p\cdot q)}^2+4\bigl (
(p^1q_1)(p^2q_2)+(p^1q_1)(p^3q_3)+(p^3q_3)(p^2q_2)\bigr )\\
 - 4 p^0 q_1 q_2 q_3 + 4q_0 p^1 p^2 p^3 
\end{equation}
and
\begin{eqnarray}
p\cdot q &=&( p^0 q_0) + ( p^1 q_1)  +( p^2 q_2)+ ( p^3 q_3)\ .
\end{eqnarray}\label{aaa}
The function $ W(p^\Lambda ,q_\Lambda)$ is symmetric under
transformations: $
p^1\leftrightarrow p^2 \leftrightarrow p^3 $
and  $ q_1\leftrightarrow q_2 \leftrightarrow q_3 $
and we have to require that $ W>0 $.
In addition to these symmetries, one can also replace each $p^\Lambda$ and $q_\Lambda$ in the expression (\ref{ww}) for $W$  by the combinations of $p^\Lambda$ and $q_\Lambda$ shown in the first column in eqs. (\ref{dictionary1}) or (\ref{dictionary2}) or (\ref{dictionary3}).

As pointed out in \cite{Duff:2006uz}, the classical expression for the entropy of the STU black holes $ W(p^\Lambda ,q_\Lambda)$ (\ref{ww})  can be represented in a very beautiful form:
\be
S^{\rm BPS}=  \pi  \sqrt W  = {\pi\over 2} \sqrt{-\rm Det ~\psi}\ , \qquad  \rm Det ~\psi  <0 \ , 
\ee
where $\rm Det ~\psi$ is the Cayley's hyperdeterminant of the unnormalized vector with components $\psi_{ijk}$ related to $p^\Lambda$ and $q_\Lambda$ by Eq. (\ref{dictionary}). The BPS black hole entropy condition $W^{\rm BPS} >0$   requires the related Cayley's hyperdeterminant to be negative.

Recently the entropy of  some examples of extremal non-BPS STU black holes have been calculated in \cite{Tripathy:2005qp}, \cite{Kallosh:2006bt}\footnote{Examples of extremal non-supersymmetric black holes were presented before in $N=8$ theory in \cite{Ortin:1996bz} where it was shown that the flip of the sign of one of the charges leads from BPS to non-BPS solution.}. We will show in \cite{K} that in general case, the entropy of non-extremal black holes in STU model is equal to 
\be
S^{\rm non-BPS}=  \pi  \sqrt {-W}= \pi \sqrt{\rm Det ~\psi}\ , \qquad  \rm Det ~\psi  >0 \ .
\ee

Thus we find that in all cases, including BPS and non-BPS, the classical supergravity entropy formula is
\be\label{generic}
S =   \pi \sqrt { | W (p,q)|}= \pi \sqrt{\rm |Det ~\psi|}= {\pi\over 2}  \sqrt{\tau_{ABC}}  \ .
\ee
Here $\tau_{ABC} = 4 |{\rm Det}~a|$ determines the three-way entanglement of the three qubits A, B and C, and  $\psi_{ijk}$ defines an unnormalized vector with the coefficients depending on $(p,q)$, see Eq. (\ref{dictionary}).

Note that because of the $[SL(2,\mathbb{Z})]^3$ invariance, the result of the calculation of the black hole entropy $S^{\rm BPS}$ does not change if instead of the hyperdeterminant of the matrix $\psi_{ijk}$ defined in Eq. (\ref{dictionary}) one uses the hyperdeterminant of the matrix $\alpha_{ijk}$ defined in (\ref{dictionary1}), or the hyperdeterminant of the matrix $\beta_{{ijk}}$ defined in (\ref{dictionary2}), or the hyperdeterminant of the matrix $\gamma_{{ijk}}$ defined in (\ref{dictionary3}).

 In string theory the microscopic entropy of the set of states with some number of branes was derived in \cite{Strominger:1996sh}  as 
\be
\log ~d(p, q) = S_{micro}(p,q) \ .
\ee
Here $d(p,q)$ counts the total number of states for a given set of integers $(p,q)$.  In the limit of large $(p,q)$ 
\be
 S_{micro}(p,q) \Rightarrow S_{macro}(p,q) \ .
\ee

For our STU model the  specific calculation was performed in \cite{Maldacena:1997de} in the context of M-theory which by duality can be related to type IIB string theory with the relation between $(p,q)$  and the numbers of $D_0, D_2, D_4, D_6$ branes shown in eq. (\ref{dictionary}). Their expression for the square of the microscopic entropy in addition to the classical expression
$\pi^{2}  W(p^\Lambda ,q_\Lambda)$,  which is  quartic in charges, contained some extra terms  quadratic in charges, which come from quantum corrections. We will come back to a more detailed discussion of these terms later.

The interest to the extremal black holes was enhanced during the last couple of years by the OSV conjecture \cite{Ooguri:2004zv} about the relation between extremal black holes and topological string theory, see for example \cite{Dabholkar:2005dt} where these recent developments are presented. In these new developments it was important to  differentiate between the so-called ``large'' and ``small'' black holes.  The  classical black hole entropy equal to 1/4 of the area of the horizon,  in the limit of very large charges when quantum corrections are small is important for defining two different kinds of extremal black holes which have analogies in definition of classes of states in ABC systems in quantum information theory.

\begin{enumerate}

 \item  Large black holes,  $S_{\rm class}   \neq 0 \quad \rightarrow \quad $  entangled GHZ class of states,  $|\rm Det ~\psi|\neq 0 $ 

  \item  Small black holes,  $S_{\rm class}   = 0  \quad \rightarrow \quad  $  non-entangled, bipartite and   W states,  $|\rm Det ~\psi=0| $
\end{enumerate}
We will present more details on GHZ canonical states and  GHZ class of states with non-vanishing 3-tangle, as well as  on non-entangled (completely separable), bipartite and   W states with vanishing 3-tangle in Sec. 5. Here we only stress the fact that these two groups are differentiated by vanishing or non-vanishing 3-tangle which coincides with the vanishing or non-vanishing  area of the horizon of the classical extremal black holes.
We used here an expression $S_{\rm class}$ to emphasize that until now we were talking about black holes without taking into account stringy quantum corrections. With account of these corrections, the classical entropy formula changes, terms quadratic in charges have to be added to the quartic expression $W$ \cite{Maldacena:1997de}. Originally there was a discrepancy between the microscopic and macroscopic entropies. After $R^{2}$ quantum corrections were included into the supergravity action in \cite{LopesCardoso:1998wt},  the discrepancy with the microscopic entropy was removed. For large black holes the extra terms provide only a small correction. However, recently a new class of extremal black holes, ``small'' black holes, was identified, for which the quantum corrections play a crucial role. It was found in \cite{Dabholkar:2004yr,Dabholkar:2004dq} that 
the ``small'' black holes  with $S_{\rm class} ={A_{\rm class}\over 4} = 0$ actually acquire a non-vanishing entropy and a non-vanishing area of the horizon, $S_{\rm quant} ={A_{\rm quant}\over 2} \neq 0$. This phenomenon is known as a ``stringy cloak for the classical singularity.''  This is a realization of the idea of a ``stretched black hole horizon'' proposed earlier by Susskind and Sen   \cite{Susskind:1993aa}, \cite{Sen:2005kj}.

Completely separable states, including, e.g., the states with only one (electric or magnetic) charge, also have a classically vanishing entropy and area of the horizon. Recently it was found in the context of the Sen's new entropy function formalism \cite{Sen:2005iz}  that the $R^{4}$ type quantum corrections  may lead to a nonvanishing entropy and stretching of the horizon even for such states \cite{Sinha:2006yy}.

%%%%%%%%%%%%%%%%%%%%

\section{Black holes, 3-qubit states and twistors}\label{twist}

The form of the STU black holes which we studied above is completely symmetric in STU variables.
This model is described by the prepotential $F={X^1 X^2 X^3\over X_0}$. The symplectic section consists of four homogeneous coordinates  $X^\Lambda$, depending on 3 special coordinates $S,\, T,\, U$, \\ $X^\Lambda=\{ X^0=1, X^1=S, X^2=T, X^3=U\}$ and four derivatives of the prepotential, $F_\Lambda \equiv {\partial F\over \partial X^\Lambda}= \{ F_0= - STU, F_1= TU, F_2= SU, F_3=ST \}$.

One can easily switch to the form in which one of the moduli is not on equal footing with others. In ABC system this would make one of the three friends, say Alice, not on equal footing with  Bob and Charlie. In black hole case we can use a symplectic transformation, a particular $Sp(8, \mathbb{Z})$ matrix, 
to transform into a new basis which has no prepotential \cite{Ceresole:1995jg}. In this new basis one of the moduli, say S, is removed from the set of new homogeneous coordinates, $\hat X^\Lambda$  and it shows up only in $\hat F_\Lambda$'s so that the total section is given by hatted coordinates $\hat X^\Lambda= {1\over \sqrt 2}\{ 1-TU, -(T+U), -(1+TU), (T-U)\}$ and $\hat F_\Lambda= S\eta_{\Lambda \Sigma} \hat X^\Sigma $.  Here $\eta_{\Lambda \Sigma}= (++--)$. The $(S|TU)$ coordinates now parametrize a coset space ${SU(1,1)\over U(1)}\times {SO(2,2)\over SO(2)\times SO(2)}$. The metric 
$\eta_{\Lambda \Sigma}= (++--)$ reflects  the manifest  $SO(2,2)$ symmetry.

In the relevant description of the ABC system one can say: Alice was promoted to the status of the 
$\hat F_\Lambda$ person whereas Bob and Charlie remain the $\hat X^\Lambda$-guys. Or, in an opposite mood one can say that Alice was excluded from the list of $\hat X^\Lambda$ persons and became an $\hat F_\Lambda$ person. Either way, she is not treated on equal footing with Bob and Charlie anymore. The corresponding transformation also produces the new hatted black hole charges $(\hat p^\Lambda, \hat q_\Lambda)$. In terms of these hatted charges our lengthy expression for the entropy given by Eqs. (\ref{ww}), (4.4)     looks very simple \cite{Behrndt:1996hu} 
\be
\rm Det ~ a= W \Big (p (\hat p, \hat q), q (\hat p, \hat q)\Big) = \hat p^2 \hat q^2- (\hat p\cdot \hat q)^2 \ .
\label{entr}\ee
Here all contractions of the hatted 4-vectors are done with the metric $\eta_{\Lambda \Sigma}= (++--)$,  $\hat p^2= \hat p^\Lambda \eta_{\Lambda \Sigma} \hat p^\Sigma=(\hat p^1)^2+(\hat p^2)^2-(\hat p^3)^2-(\hat p^4)^2$, $\hat p\cdot \hat q \equiv  \hat p^\Lambda  \hat q_\Lambda$ etc. The  duality invariant black hole entropy  described by expression in eq.  (\ref{entr}) for STU black holes was discovered in the context of $N=4$ string theory in \cite{CT2}.

The relevant 3-qubit entanglement  in this basis is given by
\be
\tau_{ABC}=4|\rm Det ~\psi|  = 4 |\hat p^2 \hat q^2- (\hat p\cdot \hat q)^2| = 2|P^{\Lambda \Sigma} P_{\Lambda \Sigma}|= |(P- *P)\cdot (P+*P)| \ ,
\ee
where the antisymmetric bivector $P^{\Lambda \Sigma}$ is defined as follows
\be
P^{\Lambda \Sigma} \equiv \hat p^{\Lambda} \hat q^{ \Sigma}- \hat p^{ \Sigma} \hat q^{\Lambda}  \ ,
\ee
where $*P$ is a dual to $P$ and $\hat q^\Lambda= \eta^{\Lambda\Sigma}\hat q_\Sigma$.

This construction may be easily compared with the description of the 3-qubit system in the context of twistor geometry \cite{Levay}. Indeed, by some operation, closely related to the change of a basis in the black hole system which requires to put e.g. Alice on non-equal status with Bob and Charlie, the  form of the 3-tangle is obtained in \cite{Levay}:
\be
\tau_{ABC}=4|\rm Det ~\psi| = 4 |(Z\cdot Z) (W\cdot W)- (Z\cdot W)^2|= 2|P^{\mu \nu} P_{\mu \nu}| \ ,
\ee
where the bivector 
\be
P^{\mu \nu} \equiv  Z^{\mu}  W^{ \nu}-  Z^{\nu} W^{ \mu} \ ,
\ee
and $Z\cdot Z= Z^\mu \eta_{\mu\nu} Z^\nu$ and $\eta_{\mu\nu}=(+++-)$, i. e.
 each vector $Z^\mu$ and $W^\mu$ is a complex vector in $SO(3.1)$ space.

Twistors associated with null vectors can be defined either in spaces with Minkowski signature $+++-$ or in spaces with $(++--)$. The relation between the corresponding 2-component spinors is the following. For the case of null vectors, $Z^\mu E_{\mu BC}= a_{0BC}$, $W^\mu E_{\mu BC}= a_{1BC}$ of \cite{Levay} one has to take the twistors  $\lambda^A_B$ and $\tilde \lambda^A_C$ in $a_{ABC}=  \lambda^A_B\tilde \lambda^A_C $ (no summation in $A$) to be related via complex conjugation, $\tilde\lambda= \pm \bar \lambda$. In $(++--)$ signature these two twistors  $\lambda_B$ and $\tilde \lambda_C$ have to be completely independent real 2-component objects, since our $SO(2,2)$ without any complexification is isomorphic to $SL(2, \mathbb{R}\times  SL(2, \mathbb{R})$. This completes the translation from the black holes in the $(S|TU)$ basis to the twistor form of the 3-qubit  ABC system in the $(A|BC)$ basis.

\begin{figure}[h!]
\centerline{ \epsfxsize 3.1 in\epsfbox{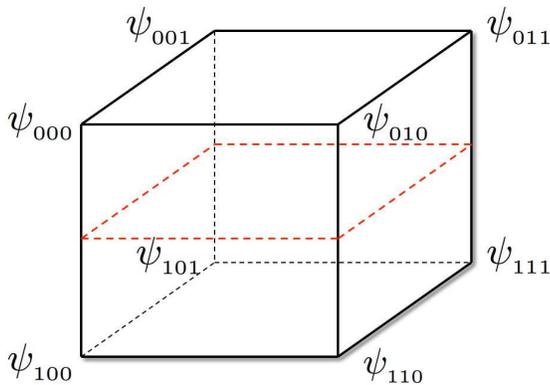}}
\caption{The $2\times 2\times 2$ matrix corresponding to twistor picture of a 3-cubit in \cite{Levay}. The combination of 4 upper corners forms a 4-vector $Z$.
All lower corners  are used to form a 4-vector $W$.}
\label{mycubesliced}
\end{figure}

To make the relation between black holes and 3-qubit states in twistor form clear, let us look at the pictures.
First, we can cut the 3-qubit cube in Fig. \ref{F1} by a horizontal surface so that all upper corners which have $0$ in the first position are used for forming a 4-vector $Z$ in \cite{Levay}. 
All lower corners, which have $1$ in the first position, are used to form a 4-vector $W$, see Fig. \ref{mycubesliced}. 

\begin{figure}[h!]
\centerline{ \epsfxsize 3.2 in\epsfbox{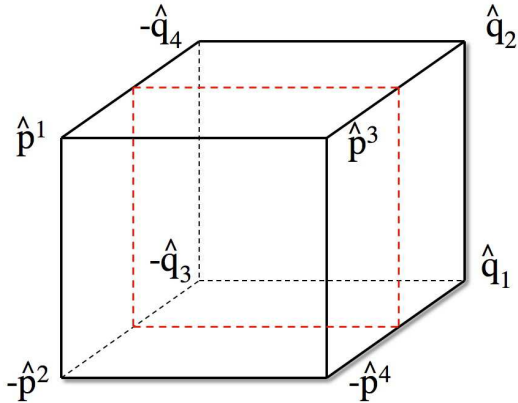}}
\caption{The $2\times 2\times 2$ matrix corresponding to supergravity black holes \cite{Behrndt:1996hu} in the hatted basis,  $\hat p$ and  $\hat q$. One has to slice this cube vertically so that the back side is cut from the front side. In this way we will separate  the 4 corners in the front forming a $\hat p$ vector and the 4 corners in the back forming a $\hat q$ vector.}
\label{stucube}
\end{figure}

 In order to see  the relation between black holes and twistors we have to use a cube which appears after an $Sp(8, \mathbb{Z})$ duality transformation to the hatted basis, see Fig. \ref{stucube}.

In the twistor formulation of the 3-qubit system, the classification of the states proceeds in simple geometric terms related to properties of the $Z^\mu$ and $W^\mu$ vectors translated into the language of the twistor theory. Using our hatted vectors  $\hat q$ and  $\hat p$ we easily perform an analogous classification for black holes. Clearly, the cube in Fig. \ref{stucube} with the vertical slice between front and back is related to the Fig. \ref{mycubesliced} after a rotation and renaming the corners.

%%%%%%%%%%%%%%%%%

\section{Classification of states of extremal black holes and  3-qubit states}\label{class}

In ABC systems there are two groups of states, each with subdivisions,   see Table \ref{Table1}, where the values of 3-tangle and local entropies are given \cite{Dur}. In group A  one finds non-entangled product space (completely separable states) and bipartite entanglement (biseparable states). In group B of
genuine entangled  3-qubit  states there are two different classes: W-class and  GHZ class. In this classification only GHZ  (Greenberger, Horne, Zeilinger) class of states \cite{Greenberger} 
corresponds to ``large'' extremal black holes (i.e. to usual extreme black holes) since $\tau_{ABC}=\Big({2S_{\rm class}\over \pi}\Big )^2 \neq 0$. 

All  states except the GHZ state (i.e. completely separable, biseparable and  W-class states) have a vanishing 3-tangle/classical entropy $\tau_{ABC}=\Big({2S_{\rm class}\over \pi}\Big )^2 = 0$.  All of these may  describe the ``small'' black holes where  ``small'' is defined by the vanishing area of the horizon of the classical black hole solution.  We will find examples of all such black holes. 

There are many ways to classify different states of the 3-qubit system. We found it most convenient to classify all possible states by discussing several ways to place charges to the corners of the cube shown in Figs. 1, 2.

\begin{table}
\centering{\begin{tabular}[t]{||c|l|l|l|l||}
\hline
~~Class~~   &~~$S_A$~~ & ~~$S_B$~~ &~~$S_C$~~& ~~$\tau_{ABC}$~~  \\ \hline
A-B-C   & ~~~0 & ~~~0 & ~~~0 & ~~~0     \\ \hline
A-BC    & ~~~0 & $>0$  & $>0$  & ~~~0     \\ \hline
B-AC    & $>0$  & ~~~0 & $>0$  & ~~~0     \\ \hline
C-AB    & $>0$  & $>0$  & ~~~0 & ~~~0     \\ \hline
W   & $>0$  & $>0$  & $>0$  & ~~~0     \\ \hline
GHZ & $>0$  & $>0$  & $>0$  & $>0$  
   \\ \hline
\end{tabular}}
\caption[]{Values of the local entropies $S_A, S_B, S_C$ defined in (\ref{locent}) and the 3-tangle $\tau_{ABC}$ for the different classes.}
\label{Table1}
\end{table}

%%%%%%%%%%%%%%%%%

\subsection{All states with vanishing 3-tangle and vanishing black hole entropy; ``small'' black holes}

For all black holes with vanishing 3-tangle $\tau_{ABC}$, i.e. with vanishing total black hole entropy, one has the following relations for the local entropies defined in (\ref{locent}):
\begin{eqnarray}
 S_A &=   {\cal C}^2_{AB}+{\cal C}^2_{AC} \ , \\
 S_B &=   {\cal C}^2_{AB}+{\cal C}^2_{BC} \ , \\
 S_C &=   {\cal C}^2_{CB}+{\cal C}^2_{AC} \ .
\label{symmetric}
\end{eqnarray}

\subsubsection{Non-entangled product space, A-B-C state.}  

\begin{figure}[h!]
\centerline{ \epsfxsize 2.7 in\epsfbox{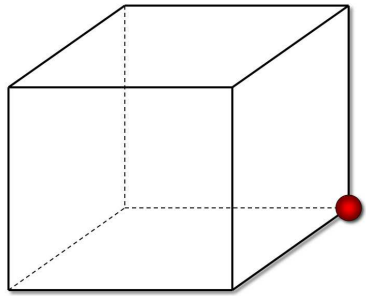}}
\caption{The $2\times 2\times 2$ matrix with all entries vanishing except one,  e.g. $q_0$. We show it by a corner with a  circle. This corresponds to a  non-entangled   completely separable state describing a black hole with just one charge,  $q_0$,  with vanishing area of the horizon. }
\label{Slide9}
\end{figure}

An easy way to see the properties of a completely separable state
  is by looking at the cube which has just one corner with a non-vanishing entry. All concurrences  are vanishing.
As an example, we may consider a black hole with just one charge, e.g. $q_0$,  with vanishing area of the horizon and null singularity,  see Fig. \ref{Slide9}. The corresponding quantum state is $|\Psi\rangle  = q_{0}|111\rangle$, i.e.  $\psi_{111} = q_0$ in the basis (\ref{dictionary}). For this state one has
\be
S_A = S_B=S_C=0\ , \qquad    {\cal C}_{AB}={\cal C}_{AC} ={\cal C}_{BC}=0\ , \qquad \tau_{ABC}=0 \ .
\ee
Quantum corrections may stretch the horizon. As a result,   this black hole may acquire a nonzero entropy proportional to $\sqrt{|q_{0}|} = \sqrt{|\Psi|}$\, \cite{Sinha:2006yy}, see Section \ref{small}. 
\begin{figure}[h!]
\centerline{ \epsfxsize 2.7 in\epsfbox{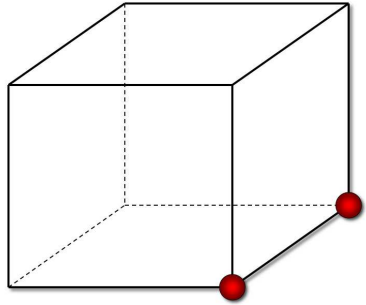}}
\caption{The $2\times 2\times 2$ matrix with two charges connected to each other by an edge. This configuration also corresponds to a  non-entangled   completely separable state describing a black hole with vanishing area of the horizon, in the classical approximation. }
\label{Slide16}
\end{figure}

One could also consider a cube with two charges connected to each other by an edge, for example, $q_0$ and $q_1$, with  $|\Psi\rangle  = q_{0}|111\rangle+ q_{1}|011\rangle$  see Fig. \ref{Slide16}. This would also represent a completely separable state; all corresponding determinants would vanish. 

It is instructive to see how the state $|\Psi\rangle  = q_{0}|111\rangle$ looks in the $S$-basis, in terms of the $S$-basis decomposition $|\Psi\rangle = \sum \alpha_{ijk}|ijk\rangle_{\alpha}$. From the dictionary Eq. (\ref{dictionary1}) one finds that $|\Psi\rangle  = q_{0}|111\rangle = a\,q_{0}|111\rangle_{\alpha} -c\, q_{0}|110\rangle_{\alpha}$. This state, up to numerical coefficients, coincides with the state shown in Fig. \ref{Slide16}. 

Similarly, when we go to the $T$-basis or $U$-basis, we will get the states $|\Psi\rangle  = a\,q_{0}|111\rangle_{\beta} -c\, q_{0}|101\rangle_{\beta}$ and  $|\Psi\rangle  = a\,q_{0}|111\rangle_{\gamma} -c\, q_{0}|011\rangle_{\gamma}$. In all of these cases we obtain states described by the cubes with the charge $a\,q_{0}$ in the same position as in Fig. \ref{Slide16} and with a second charge $-c\, q_{i}$ connected to it by an edge.  All of these cases belong to the same class of completely separable states.

If one tries to add more charges, or place them differently (i.e. add charges  $p^{\Lambda}$ to an already existing charge $q^{\Lambda}$), one can only produce states that will not be completely separable. Therefore the simple cube with one entry, Fig. \ref{Slide9}, represents the general class of all completely separable states.

\subsubsection{Bipartite entanglement;   A-BC state.} 
\begin{figure}[h!]
\centerline{ \epsfxsize 2.7 in\epsfbox{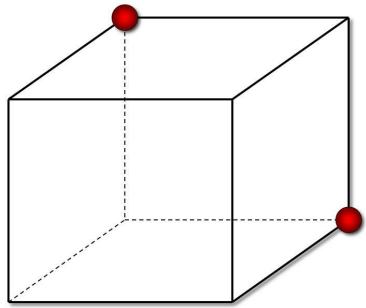}}
\caption{The $2\times 2\times 2$ matrix with all entries vanishing except two entries on the same side but in opposite corners. They are shown by  circles, one for $\psi_{111}=q_0$ and one for $\psi_{001}=-p^1$. This is the case of the ``small'' black hole with just 2 charges $q_0$ and $p^1$ and with classically  vanishing area of the horizon. }
\label{Slide11}
\end{figure}

 In order to obtain a biseparable state one may consider a cube with two non-vanishing entries in the opposite corners  of one side of the cube so that there is one non-vanishing 2-tangle, for example $\tau_{BC}$. This state shown in Fig. \ref{Slide11} describes a black hole with charges $q_{0}$ and $p^{1}$, which corresponds to a quantum state  $
|\Psi\rangle  = -p^{1}|001\rangle+q_{0}|111\rangle$ (the signs are due to the translation between the charges and $\psi_{ijk}$, Eq. (\ref{dictionary})). In this case we have two non-vanishing  entanglements between the 1-qubit  and a 2-qubit system (or 2 local entropies). 
\begin{eqnarray}
 S_A &=& 0\ , \\
 S_B &=&{\cal C}_{BC}^2=  4|q_0 p^1|^2   \neq 0 \ , \\
 S_C &=& {\cal C}_{BC}^2  =4|q_0 p^1|^2 \neq 0  \ .
\label{sep}
\end{eqnarray}

This is the ``small'' black hole with just 2 charges and with classically  vanishing entropy and the area of the horizon \cite{Dabholkar:2004dq}.  When quantum corrections are included, which lead to quantum stretching of the horizon, the value of the new area is proportional to the only non-vanishing concurrence  of the  BC system inside the ABC system, ${\cal C}_{BC}=2|q_0 p^1|$, see Section \ref{small}. It  is also a concurrence 
${\cal C}_{B(AC)}=2|q_0 p^1|$ between Bob and the system of Alice-Charlie as well as a concurrence ${\cal C}_{C(AB)}=2|q_0 p^1|$ between Charlie and the system of Alice-Bob.

\subsubsection{W-class of states }

\begin{figure}[h!]
\centerline{ \epsfxsize 2.7 in\epsfbox{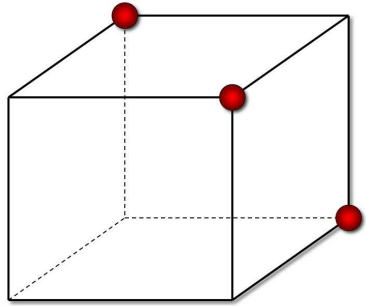}}
\caption{The $2\times 2\times 2$ matrix with all entries vanishing except for $\psi_{111}=q_0$, $\psi_{001}=-p^1$ and $\psi_{010}=-p^2$, corresponding to the charges $q_0$, $p^1$ and $p^2$.   The charges are always in opposite corners of each of these 3 sides. This state describes the ``small'' black hole \cite{Dabholkar:2004dq} with just 3 charges and with classically  vanishing area of the horizon.}
\label{Slide12}
\end{figure}

Now let us consider 3 entries in the black hole case: $q_0$, $p^1$ and $p^2$ charges, as shown in Fig. \ref{Slide12}. This is the state $
|\Psi\rangle  =- p^{1}|001\rangle-p^{2}|010\rangle+q_{0}|111\rangle$.  None of the local entropies is vanishing, however we still have a vanishing 3-tangle and, at the classical level, vanishing entropy and the area of the horizon \cite{Dabholkar:2004dq}.  The corresponding black holes may be corrected and the area of the horizon with account of quantum corrections may depend on  $q_0 p^1$ and $q_0 p^2$. Here again we will find that the stretched horizon depends on  non-vanishing concurrences of the 2-qubit systems inside ABC,  see Section \ref{small}.
\begin{eqnarray}
 {\cal C}_{AB} &=    2|q_0 p^1| \neq 0\ ,\\
  {\cal C}_{BC} &= 2|q_0 p^2| \neq 0\ ,\\
  {\cal C}_{AC}  & = 2|p^1 p^2| \neq 0 \ .
\label{symmetric2}
\end{eqnarray}
and
\begin{eqnarray}
 S_A &=&{\cal C}^2_{AB}+{\cal C}^2_{AC} =   (p^1)^2\Big ( (q_0)^2 + (p^2)^2 \Big )\ , \\
 S_B &=&{\cal C}^2_{AB}+{\cal C}^2_{BC}=(q_0)^2 \Big( (p^1)^2 + (p^1)^2 \Big) \ , \\
 S_C &=&{\cal C}^2_{CB}+{\cal C}^2_{AC}=(p^2)^2 \Big ( (q_0)^2 + (p^1)^2 \Big) \ .
\label{symmetric3a}
\end{eqnarray}

%%%%%%%%%%%%%%%%

\subsection{Non-vanishing 3-tangle and entropy, GHZ states; ``large'' black holes}

\begin{figure}[h!]
\centerline{ \epsfxsize 2.7 in\epsfbox{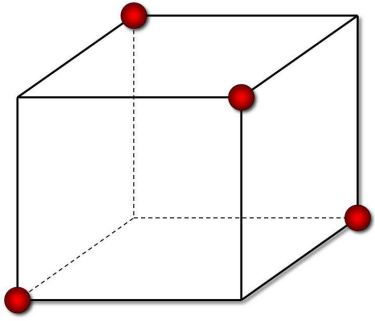}}
\caption{The $2\times 2\times 2$ matrix with 4 non-vanishing charges, for example, $q_0$, $p^1$,  $p^2$, $p^3$. This is  a case of the ``large'' BPS and non-BPS black holes (depending on the sign of the product of these four charges) with just 4 charges and with classically  nonvanishing area of the horizon. It belongs to the GHZ class of  states, which may describe either BPS or non-BPS black holes.}
\label{Slide13}
\end{figure}

Here we have to satisfy the equations (\ref{3tangleA})-(\ref{3tangleC})
with non-vanishing left hand side.
Using our cube pictures, we may immediately see that the configuration in Fig. \ref{Slide13} corresponds to a class of GHZ states, where we pick up some set of black hole charges in the expression for the non-vanishing entropy. For example, in the case of  supersymmetric BPS black holes we may have non-vanishing charges $q_0, p^1, p^2, p^3$ with $q_0\, p^1\, p^2\, p^3 > 0$. We place them as shown in Fig. \ref{stu2cube}. This is the cube in Fig. \ref{Slide13}. The corresponding quantum  state is $
|\Psi\rangle  =- p^{1}|001\rangle-p^{2}|010\rangle- p^{3}100\rangle+ q_{0}|111\rangle$. Now every side has two non-vanishing entries so that a concurrence associated with each side is non-vanishing. More importantly, the entropy and the 3-tangle also do not vanish, 
\be\label{genericnew}
S =   \pi \sqrt { | W (p,q)|}= \pi \sqrt{\rm |Det ~\psi|}= {\pi\over 2}  \sqrt{\tau_{ABC}}  = 2\pi\sqrt{|q_0\, p^1\, p^2\, p^3|} \ .
\ee

If however, $q_0 p^1 p^2p^3 <0$, this will be related to an extremal non-supersymmetric non-BPS black hole with 4 charges,  \cite{Tripathy:2005qp}, \cite{Kallosh:2006bt}. This is in general the case when $W<0$ \cite{K}. 

\begin{figure}[h!]
\centerline{ \epsfxsize 2.7 in\epsfbox{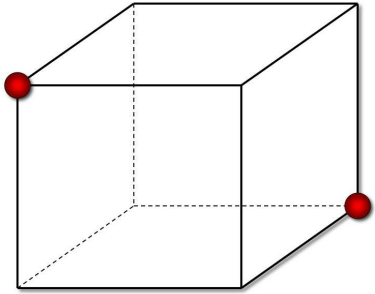}}
\caption{The $2\times 2\times 2$ matrix with all entries vanishing, but two on the  opposite diagonal of the cub.  This is  a case of the ``large'' non-BPS black hole with just 2 charges (in one gauge group, like $p^0$ and $q_0$)  and with classically  nonvanishing area of the horizon. It corresponds to the canonical GHZ  state describing non-BPS black holes.}
\label{Slide10}
\end{figure}

By using transformations preserving $\tau_{ABC}= |{\rm Det} ~\psi|$ (but not necessarily the sign of ${\rm Det} ~\psi$) one can always transform a state  $
|\Psi\rangle  =- p^{1}|001\rangle-p^{2}|010\rangle -p^{3}100\rangle+ q_{0}|111\rangle$ to a canonical GHZ state describing only one electric and one magnetic charge in the same gauge group, say $\tilde p^0$ and $\tilde q_0$:~  $
|\Psi\rangle  = \tilde p^{0}|000\rangle+ \tilde q_{0}|111\rangle$, see  Fig. \ref{Slide10}. The two charges corresponding to a canonical GHZ state are always at the opposite corners of the cube.  One can easily check that for the canonical GHZ states, Fig. \ref{Slide10}, the Cayley's hyperdeterminant ${\rm Det} ~\psi$ is always positive and $W$ is always negative, which corresponds to non-supersymmetric non-BPS black holes.

Thus all extremal BPS and non-BPS black holes with non-vanishing entropy, i.e. all ``usual,'' or ``large'' black holes, belong to the GHZ class of states of the ABC system, which is described by the lowest line in Table 1. However, the theory of stringy black holes requires a more detailed classification than the standard 3-qubit classification provided by Table 1. One encounters  two inequivalent subclasses of  GHZ states with respect to supersymmetry. The canonical GHZ states, Fig. \ref{Slide10}  always  correspond to non-supersymmetric non-BPS black holes. Meanwhile the GHZ states described by Fig. \ref{Slide13} have the same 3-tangle (i.e. the same $|{\rm Det} ~\psi|$), but the sign of ${\rm Det} ~a$ may be either positive or negative. The states with ${\rm Det} ~\psi > 0$ correspond to non-supersymmetric non-BPS black holes, whereas the states with ${\rm Det} ~\psi< 0$ correspond to supersymmetric BPS black holes.

%%%%%%%%%%%%%%%%%%%%%%

\section{Entropy of ``small'' black holes, the norm and 2-tangles in ABC systems}\label{small}

One of the goals of our paper was to obtain a better understanding of the intriguing relation between the entropy of the extreme BPS STU black holes and the 3-tangle discovered by Duff \cite{Duff:2006uz}. In this paper we extended his analysis for the axion-dilaton black holes and for the non-BPS STU black holes, and  developed a new set of tools for investigation and classification of black holes, which have their counterparts in the theory of quantum information. Now we may  apply our tools to the so-called extremal ``small'' black holes, which have a singular horizon with vanishing area and zero entropy at the classical level, but may acquire nonvanishing entropy and the area of horizon due to quantum corrections. 

Let us first consider completely separable states such as a state with a single charge $q_{0}$ shown in Fig. \ref{Slide9}.
The corresponding ``small'' black holes with just one charge (e. g. number of D0 branes) were studied in \cite{Sinha:2006yy}. The value of the entropy due to $R^4$ corrections in the limit  $q_{0} \gg 1$ was  found to be 
\be
S_{BH}= \pi K\sqrt {{2\over 3} |q_{0}|} \ , 
\label{D0}\ee
where $K $ is some number. This entropy is also proportional to the area of the stretched horizon. As emphasized in \cite{Sinha:2006yy}, in order to verify this result one may need to check higher order corrections in $R$. If Eq. (\ref{D0}) is valid, one can represent it in the form that does not depend on the choice of a single charge  $q_{\Lambda}$ or $p^{\Lambda}$:
\be\label{comsep}
S_{BH}= K \sqrt {{2\over 3} |\Psi|}   \ ,
\ee
 where $|\Psi|$ is the norm of the state defined in eq. (\ref{|Psi|}). One can interpret this result as a consequence of the quantum stretching of the horizon conjectured by Susskind and Sen \cite{Susskind:1993aa}.   The classification in Table 1 does not attach any invariant concept to completely separable states, simply because in the quantum information theory all of these states are equally normalized: $|\Psi| = 1$. Meanwhile  the entropy of the black holes is proportional to the square root of the  wave function with a ``stretched'' norm $|\Psi| =|q_0| \not = 1$ (\ref{|Psi|}). Thus we arrive at a simple intuitive interpretation: the stretching of the horizon of black holes with a single charge is related to the stretching of the norm of this state $|\Psi|$.

Now let us consider the bipartite case characterized by the charges $q_0,\, p^1 $, numbers of D0 and D4 branes. In this case, the entropy with account of quantum corrections calculated in supergravity  is given by 
\be
S_{\rm quant} =  4 \pi \sqrt {|q_0 p^1| } \ . 
\ee
This entropy  was calculated in  \cite{Dabholkar:2004yr} by counting the number of microstates of string theory for $q_0\gg  p^1\gg 1 $.  By comparing this answer with  Fig. \ref{Slide11}, one can see that the only parameter available in the classical cube is precisely the non-vanishing concurrence, ${\cal C}=2|q_0 p^1|$,  so we have the following interpretation of this result:
\be
S_{\rm quant}  =  \pi \sqrt {{\cal C}} \ . 
\ee
The radius of the stretched horizon $r_{h}$ and the area of the horizon of the ``small'' black holes $A_{\rm quant} = 2S_{\rm quant}$ were calculated in \cite{Dabholkar:2004dq}. Now we see that they have the following  interpretation in terms of the concurrence of the 2-cubit state inside a 3-qubit state in quantum information theory:  
\be
{A_{\rm quant}} =  4\pi r_h^2 = 2 \pi \sqrt {\cal C} \ .
\ee

It is amazing that   the quantum corrected area of the horizon and entropy are related to the only non-vanishing concurrence for the case of the bipartite state $q_0,\, p^1 $. One may wonder how quantum corrections in string theory could know about the concurrences in 3-qubit systems? Is it just another coincidence or simply a consequence of the underlying symmetry of the theory?

 Now let us make another step and discuss the entropy and the area of the horizon of the black holes in the bipartite or W-state  with non-vanishing  $q_0, p^1, p^2$ charges. At the classical level, such black holes have a vanishing singular horizon with null singularity and zero entropy. Meanwhile quantum effects 
give the entropy  \cite{Dabholkar:2004dq}
\be\label{twocharges}
S_{\rm quant}  = 4 \pi \sqrt {|q_0 (p^1+ p^2)|} = {A_{\rm quant}\over 2} \ .
\ee
These calculations, and  the semi-classical approximation in general,  require the condition that $q_0\gg p^1, p^2 \gg 1$. In such case the term $p^1 p^2$, which is naturally expected from the cube picture, may be missing simply because it is supposed to be much smaller than the other two terms.  In the limit $q_0\gg p^1, p^2 \gg 1$ one can describe all results concerning the entropy of ``large'' and ``small'' black holes in the bipartite or W-state by one simple equation preserving the symmetries of the system: 
\be\label{totentr}
S_{\rm total} =  {\pi\over 2} \sqrt { \tau_{ABC} +{4 \, c_2\over 3 }({\cal C}_{AB} + {\cal C}_{BC}+{\cal C}_{CA})} \ .
 \ee
Here $c_2$ is the second Chern class coefficient of the compactified manifold; in the example of K3 manifold $c_2=24$. Interestingly, ${\cal C}_{AB} + {\cal C}_{BC}+{\cal C}_{CA}$ is equal to a half of the total area of a box with sides $|q_{0}|$, $|p^1|$ and $|p^2|$. 
The total entropy has two contributions, $\tau_{ABC}$, which is quartic in charges, and  ${4 \, c_2\over 3}({\cal C}_{AB} + {\cal C}_{BC}+{\cal C}_{CA})$, which is quadratic in charges.
Therefore for ``large'' black holes this expression in the leading approximation agrees with the result obtained in \cite{Behrndt:1996hu} and coincides with the result obtained by counting of states in string theory \cite{Maldacena:1997de} and in supergravity  with $R^{2 }$ corrections \cite{LopesCardoso:1998wt} under the condition that $q_0\gg p^1, p^2 \gg 1$. For  ``small'' black holes  the classical entropy vanishes, $\tau_{ABC}=0$, and the microscopic entropy  calculated in string theory \cite{Dabholkar:2004yr,Dabholkar:2004dq} is reproduced correctly by Eq. (\ref{totentr}) in the approximation $q_0\gg p^1, p^2 \gg 1$:
\be
S_{\rm
small} = {A_{\rm
small}\over 2} = \pi \sqrt { { \, c_2\over 3 }({\cal C}_{AB} + {\cal C}_{BC}+{\cal C}_{CA})} \ . \ee

One may go one step further and consider the ``small'' 1-charge black holes \cite{Sinha:2006yy}. The modified entropy formula, under the conditions specified above, can be written as follows:
\be\label{totentr1}
S_{\rm total} =  {\pi\over 2} \sqrt { \tau_{ABC} +{4 \, c_2\over 3 }({\cal C}_{AB} + {\cal C}_{BC}+{\cal C}_{CA}) + {8 K^2\over 3} |\Psi|} \ .
 \ee
Here $|\Psi|=  \sqrt {\langle \Psi| \Psi \rangle}$ is the norm of the wave function.

One can understand this equation as follows. For  completely separable states with only one nonzero charge, this equation  is reduced to Eqs. (\ref{comsep}), (\ref{D0}). For the bipartite and W-states at large values of charges, the concurrences are much greater than $|\Psi|$, and equation is reduced to  (\ref{totentr}), which is equivalent to equation (\ref{twocharges}) in the region of its applicability. Finally, for the GHZ states the 3-tangle is much greater than the concurrences, and we return to the equation $S  =  {\pi\over 2} \sqrt { \tau_{ABC}}$. 

One may think that if eq. (\ref{totentr1}) is correct beyond just representing various known cases, in more general situation it may be a prediction of certain sub-leading corrections. Are these corrections large enough to be measured in gravity with account of string corrections, higher order curvature corrections in supergravity? The relevant results on subleading corrections to the entropy of black holes with the classically finite horizon area were derived in \cite{LopesCardoso:1998wt} and may be related to eq. (\ref{totentr1}). One can also try to relate it to the black holes studied with the tools of topological string theory in \cite{Ooguri:2004zv}.

%%%%%%%%%%%%%%%%%%%%%%%%%%

\section{$E_{7(7)}$ quartic invariant and Cayley's hyperdeterminant}\label{e77}

In the previous investigation we mostly discussed axion-dilaton black holes and STU black holes. This  covers a very broad class of extreme stringy black hole solutions.
The STU black holes are described by 8 parameters and the classical entropy of these black holes is given by a square root of the absolute value of the Cayley's hyperdeterminant. 

Now we are going to significantly generalize our results. The most general class of black holes in N=8 supergravity/M-theory is defined by 56 charges, and the entropy  formula is given by the square root of the quartic Cartan-Cremmer-Julia $E_{7(7)}$ invariant \cite{Kallosh:1996uy}, \cite{Cartan}-\cite{Ferrara:1997uz},
\be\label{kol}
 S = \pi \sqrt{|J_{4}|} \ ,
\ee
where the Cartan-Cremmer-Julia form of the invariant \cite{Cremmer:1979up}  depends on the central charge matrix $Z$,
\begin{eqnarray}
J_{4} &=& + {\mbox Tr} (Z\bar Z )^2 -{\textstyle{1\over 4}}
({\mbox Tr}\,Z\bar Z )^2 + 4  ({\mbox P\hskip- .1cm f}\; Z +
{\mbox P\hskip- .1cm f} \; \bar Z \,)  \ ,
\label{diamond}
\end{eqnarray}
and the Cartan form \cite{Cartan} depends on the quantized charge matrix $(x, y)$
\begin{eqnarray}
J_{4}&=&-{\mbox Tr} (\, x \; y )^2 + {\textstyle{1\over 4}}
({\mbox Tr}\, x \;   y)^2 - 4  ({\mbox P\hskip- .1cm f}~ x +
{\mbox P\hskip- .1cm f} ~ y \,)  \ .
\label{cartan}\end{eqnarray}
Here
\begin{equation}
Z_{AB} = -{1\over 4\sqrt 2}(x^{ab} + i y_{ab})(\Gamma^{ab})_{AB}
\label{Z}\end{equation}
is the central charge matrix and
\begin{equation}
x^{ab} + i y_{ab} = -{\sqrt 2\over 4} Z_{AB}  (\Gamma^{AB})_{ab}
\label{xy}\end{equation}
is a matrix of the quantized charges related to some numbers of branes.
The exact relation between the Cartan invariant  in eq. (\ref{cartan})  and Cremmer-Julia  invariant \cite{Cremmer:1979up} in eq. (\ref{diamond}) has been established in \cite{Balasubramanian:1997az}.

The matrices of $SO(8)$ algebra are $(\Gamma^{ab})_{AB}$ where $(a \, b)$ are the 8 vector indices and $(A,B)$ are the 8 spinor indices. The $(\Gamma^{ab})_{AB}$ matrices can be considered also as $(\Gamma^{AB})_{ab}$ matrices due to equivalence of the vector and spinor representations of the $SO(8)$ Lie algebra. The central charge matrix $Z_{AB}$ can be brought to the canonical basis for the skew-symmetric matrix using an $SU(8)$ transformation. The eigenvalues $z_i, i=1,2,3,4$ are  complex. In this way the content of a theory is reduced from 56 entries to 8.

Relation between the entropy of stringy black holes and the Cartan-Cremmer-Julia $E_{7(7)}$ invariant was established 10 years  ago \cite{Kallosh:1996uy}.   The stringy  solutions  in $N=4$ theory characterized by 5 parameters were first found in \cite{CT2}. 
Since that time many new black hole solutions have been found.
In a systematic treatment in \cite{FM}  in the context of the eigenvalues of the central charge matrix of $N=8$ theory the meaning of these 5 parameters was clarified: $z_i=\rho_i e^{i\phi}$,  from 4 complex values of $z_i=\rho_i e^{i\phi_i}$ one can remove 3 phases  by an $SU(8)$ rotation, but the overall phase cannot be removed. Therefore a 5-parameter solution is called a generating solution for other black holes in N=8 supergravity/M-theory.
 Expression for their entropy is always given by   $S = \pi \sqrt{|J_{4}|}$ for some subset of 5 of the 8 parameters mentioned above. Recently a new class of solutions was discovered, describing black rings. The maximal number of parameters for the known solutions  is 7. The entropy of black ring solutions found so far was identified in \cite{Bena:2004tk} with the expression for $\pi \sqrt{|J_{4}|}$ for a  subset of 7 out of 8 parameters mentioned above. That is why it would be most interesting to establish a possible relation between the general black hole/black ring entropy equation $ S = \pi \sqrt{|J_{4}|}$ in N = 8 supergravity/M-theory and some of the constructions of the theory of quantum information.

One could expect that this relation, if possible at all, may be quite involved and may require investigation of more complicated constructions, such as $n$-tangles for $n > 3$. However, we have found that this relation again involves only 3-tangles.

To find this relation, let us note that 
in $x, y$ basis only $SO(8)$ symmetry is manifest, which means that every term in eq. (\ref{cartan}) is invariant only under $SO(8)$ symmetry. However, is was proved in \cite{Cartan} and  \cite{Cremmer:1979up} that the sum of all terms in eq. (\ref{cartan}) is invariant under the full $SU(8)$ symmetry, which acts as follows
\be
\delta(x^{ab} \pm i y_{ab})= (2\Lambda^{[ a}{}_{[c} \delta^{b]}{}_{d]} \pm i\Sigma_{abcd}) (x^{cd} \mp i y_{cd}) \ .
\ee
The total number of parameters is 63, where 28 are from the manifest $SO(8)$ and 35 from the antisymmetric self-dual $\Sigma_{abcd}= {}^*\Sigma^{abcd} $. Thus one can use the $SU(8)$ transformation of the complex matrix $x^{ab} + i y_{ab}$ and bring it to the canonical form with some complex eigenvalues $\lambda_I,  I=1,2,3,4$. The value of the quartic invariant (\ref{cartan}) will not change. 
\be
(x^{ab} + i y_{ab})_{\rm can}= 
\begin{pmatrix}
  0 & \lambda_{1} & 0 & 0 & 0 & 0 & 0 & 0 \\
-\lambda_{1} & 0 & 0 & 0 & 0 & 0 & 0 & 0 \\
0 & 0 & 0 & \lambda_{2} & 0 & 0 & 0 & 0 \\
0 & 0 & -\lambda_{2} &0 & 0 & 0 & 0 & 0 \\
0 & 0 & 0 & 0 & 0 & \lambda_{3} & 0 & 0 \\
0 & 0 & 0 & 0 & -\lambda_{3} & 0 & 0 & 0 \\
0 & 0 & 0 & 0 & 0 & 0 & 0 & \lambda_{4} \\
0 & 0 & 0 & 0 & 0 & 0 & -\lambda_{4} & 0\end{pmatrix}
\label{Zcan}\ee
One can easily check that the  Cartan-Cremmer-Julia quartic invariant $J_{4}$ depending on 4 complex eigenvalues $\lambda_I$ can be represented as a  Cayley's hyperdeterminant of a matrix $\psi_{ijk}$
\be
J_{4}(\lambda)=  - \rm Det~ \psi \ ,
\ee
where the relation between the complex coefficients $\lambda_{i}$, the parameters $x_{ij}$ and $y^{kl}$, the matrix $\psi_{ijk}$ and the black hole charges $p^{i}$ and $q_{k}$ is given by the following dictionary:
\bea\label{dictionary5}
\lambda_{1}~ = & x_{12}+i y^{12}~ =& a_{111}+i a_{000}~ = ~~q_0+i p^0\nonumber\\
\lambda_{2}~ =& x_{34}+i y^{34}~ =& a_{001}+ i a_{110}~ =-p^1+iq_1  \nonumber\\
\lambda_{3}~ =& x_{56}+i y^{56}~ =& a_{010}+i a_{101}~ =-p^2+iq_2\nonumber\\
\lambda_{4}~ =& x_{78}+i y^{78}~ =& a_{100}+i a_{011}~ =-p^3+iq_3 
\end{eqnarray}

The simplest way to prove it is to write the quartic $E_{7(7)}$ Cartan invariant  in the canonical basis $(x_{ij},y^{ij})$, $i,j =1,...,8$:
\bea\label{quarticshort}
 J_4 &= -(x_{12}y^{12} + x_{34}y^{34}
+x_{56}y^{56}+x_{78}y^{78})^2-
4(x_{12}x_{34}x_{56}x_{78}+y^{12}y^{34}y^{56}y^{78})\cr & +
4(x_{12}x_{34}y^{12}y^{34}+ x_{12}x_{56}y^{12}y^{56} +
x_{34}x_{56}y^{34}y^{56}+x_{12}x_{78}y^{12}y^{78}+ x_{34}x_{78}
y^{34}y^{78}\cr &+x_{56}x_{78}y^{56}y^{78})\ . 
\eea
Then one should compare it to the Cayley's hyperdeterminant (\ref{cayley}) using the dictionary (\ref{dictionary5}) given above, or an equivalent dictionary in the form similar to the one used in Section  \ref{stu3}, Eq. (\ref{dictionary}):
\be
\begin{array}{c}
ABC\\
\\
\left(
\begin{array}{c}
a_{000}\\
a_{001}\\
a_{010}\\
a_{100}\\
a_{111}\\
a_{110}\\
a_{101}\\
a_{011}
\end{array}
\right)
\end{array}
~~~~~=
\begin{array}{c}
STU~Black~Hole\\
\\
\left(
\begin{array}{c}
~~p^{0}\\
-p^{1}\\
-p^{2}\\
-p^{3}\\
~~q_{0}\\
~~q_{1}\\
~~q_{2}\\
~~q_{3}
\end{array}
\right)
\end{array}
~~~=
\begin{array}{c}
N = 8~Black~Hole\\
\\
\left(
\begin{array}{c}
y^{12}\\
x_{34}\\
x_{56}\\
x_{78}\\
x_{12}\\
y^{34}\\
y^{56}\\
y^{78}
\end{array}
\right)
\end{array}
\label{dictionary6}\ee

Our results imply that the entropy of the most general extremal BPS and non-BPS black hole and black ring solutions in $N = 8$ supergravity can be brought to a  canonical  basis where it depends only on 8 charges and can be represented by the same  compact expression   (\ref{generic}) as in the theory of STU black holes and as a 3-tangle in a 3-qubit system:
\be
 S_{(BH, BR)} = \pi \sqrt{|J_{4}(\lambda)|} = \pi  \sqrt {|\rm Det~ \psi|} = \pi  \sqrt {| W (p,q)|}  = {\pi\over 2} \sqrt{\tau_{ABC}}\ .
\label{main}\ee

The quartic invariant of the $E_{7(7)}$ $J_4$ is related to the octonionic Jordan algebra $J_3^{\mathbb{O}}$, see \cite{Ferrara:1997uz}. It is therefore natural, in view of our result (\ref{main}), to expect that the 3-qubit system can be described by octonions, which was indeed  shown in \cite{Bernewig}.

\section{Conclusions}

Our work, following the recent work by Duff  \cite{Duff:2006uz}, demonstrated a lot of intriguing connections between extremal black holes and the ABC system in the quantum information theory. The new approach to the theory of stringy black holes  may help us with the black hole and black ring classification and with interpretation of our results in terms of general quantum mechanical systems. It may also help us to represent our results in a different form, which may allow our intuition to grow in a previously unexpected way.   In this paper we found that  the entropy of the axion-dilaton extremal black hole is related to the concurrence of a 2-qubit state, whereas the entropy of the STU black holes, even if they are not BPS black holes,  is related to the 3-tangle of a 3-qubit state.  We identified usual  black holes with the maximally entangled GHZ-class of states, and  ``small''  black holes with either separable, or bipartite entangled states or W-class of states. We established a certain relation between 3-qubit states,  twistors and  black holes. We found an expression for entropy and the area of the horizon of ``small'' black holes in terms of the concurrence of the 2-cubit states inside a 3-qubit state and its norm.  Finally, we extended the previous results to the most general extremal BPS and non-BPS black hole and black ring solutions in $N = 8$ supergravity/M-theory. To our own surprise, we have found that the expression for the entropy of these solutions in terms of the quartic $E_{7(7)}$ Cartan invariant \cite{Kallosh:1996uy} in eq. (\ref{cartan}) can be  represented by the same  compact expression  in terms of the Cayley's hyperdeterminant (\ref{cayley})  as a 3-tangle (\ref{coffm}) and the entropy of STU black holes  (\ref{generic}).

Our work was devoted to the implications of the quantum information theory to the theory of black holes. Even if some of these results eventually will be interpreted as coincidental, we may still appreciate  the theory of quantum information for its heuristic potential, which allowed us to look at the theory of stringy  black holes from a completely different perspective. However, we do not think that this is a one-way road. It is quite plausible that the enormous amount of highly nontrivial results obtained in the quantum theory of stringy black holes may lead to new insights in the theory of quantum information. We hope therefore that the parallel study of both sides of the story may be quite fruitful.

\newpage

{\bf Note added}

Recently a paper by Levay \cite{Levay:2006kf} appeared where two important developments were made.
First, it was shown that a pure three-qubit state is real under certain conditions. Note that in  general in  QIT the wave function is complex, and the system has a $[SL(2,\mathbb{C})]^3$ symmetry, whereas for black holes we have only $[SL(2,\mathbb{Z})]^3$ and the relevant ``wave function'' is real. Interestingly, two different conditions for reality found in \cite{Levay:2006kf} correspond to either BPS or non-BPS black holes. Secondly, it was established there that what in string theory is known as a stabilization of moduli near the black hole horizon in QIT is known as a procedure of finding the optimal local distillation protocol of a GHZ state from an arbitrary three-qubit state.
Both statements suggest that indeed it may be useful to study both sides of the story.

%\
 
{\Large {\bf Acknowledgments}}

It is a pleasure to thank T. Banks, M. Duff, S. Ferrara, A. Guth, C. Hull, S. Kachru, A. Maloney, J. Park, S. Shenker, M. Shmakova,  A. Sinha, L. Susskind and W. Taylor
for useful conversations. This work was supported by
NSF grant PHY-0244728.

\end{document}